\documentclass[preprints,review,accept,moreauthors,dvi2pdf]{mdpi} 

\renewcommand{\linenumbers}{}

\firstpage{1} 
\pubvolume{1}
\issuenum{1}
\articlenumber{0}
\pubyear{2021}
\copyrightyear{2020}
\datereceived{} 
\dateaccepted{} 
\datepublished{} 
\hreflink{https://doi.org/} 
\Title{Fuzzy instantons in landscape and swampland: review of the Hartle-Hawking wave function and several applications}

\Author{Dong-han Yeom $^{1,2}$\orcidA{}*}

\AuthorNames{Dong-han Yeom}

\AuthorCitation{Yeom, D.}

\address{%
$^{1}$ \quad Department of Physics Education, Pusan National University, Busan 46241, Republic of Korea\\
$^{2}$ \quad Research Center for Dielectric and Advanced Matter Physics, Pusan National University, Busan 46241, Republic of Korea
}

\corres{Correspondence: innocent.yeom@gmail.com}

\abstract{The Euclidean path integral is well approximated by instantons. If instantons are dynamical, then instantons are necessarily complexified. These fuzzy instantons can have various physical applications. In slow-roll inflation models, fuzzy instantons can explain the probability distribution of the initial conditions of the universe. Even though the potential shape does not satisfy the slow-roll conditions following the swampland criteria, still the fuzzy instantons can explain the origin of the universe. If we extend the Euclidean path integral beyond the no-boundary proposal, we may study fuzzy Euclidean wormholes that have various physical applications in cosmology and black hole physics. We summarize them and discuss possible future research topics.}


\begin{document}
\setcounter{section}{-1} 

\section{Introduction: preliminaries}

Understanding the nature of the origin of the Universe is one of the most fundamental problems in modern physics. Due to the singularity theorem \cite{Hawking:1969sw}, if we move backward in time, assuming very reasonable physical conditions, we expect that there must exist the initial singularity. At this singularity, all tools of general relativity break down, and hence, we need a quantum gravitational prescription.

To understand the initial singularity, the quantum gravitational description must be non-perturbative. The most conservative approach is to quantize the gravitational degrees of freedom following the canonical quantization method \cite{DeWitt:1967yk}. After applying this approach, one can obtain the quantized Hamiltonian constraint, or so-called the Wheeler-DeWitt equation. If we solve the equation, in principle, we can obtain the probability for a given hypersurface and corresponding field configurations.

One of the problems of the canonical quantization is that the probability depends on the choice of the boundary conditions \cite{Vilenkin:1987kf}. By choosing a boundary condition, one may or may not provide a reasonable probability distribution for the early universe. Of course, there is no fundamental principle to choose the boundary condition and, in principle, the boundary condition must be confirmed by possible observational consequences \cite{Halliwell:1984eu}.

\subsection{Hartle-Hawking wave function}

Now we ask what is the most natural assumption for the boundary condition of the universe. Regarding this, one may think that the ground state wave function may correspond to the most natural choice for the boundary condition. One potential problem is that the ground state is not well defined in the context of quantum gravity. However, one may reasonably argue that the Euclidean path integral might be the ground state wave function of the Wheeler-DeWitt equation \cite{Hartle:1983ai}. The mathematical form, so-called the \textit{Hartle-Hawking wave function}, is as follows:
\begin{eqnarray}
\Psi\left[h_{\mu\nu},\psi\right] = \int \mathcal{D}g_{\mu\nu}\mathcal{D}\phi \;\; e^{-S_{\mathrm{E}}\left[g_{\mu\nu},\phi\right]},
\end{eqnarray}
where $g_{\mu\nu}$ is the metric, $\phi$ is a matter field, $S_{\mathrm{E}}$ is the Euclidean action, and we sum over all regular and compact Euclidean geometries and field configurations satisfying conditions $\partial g_{\mu\nu} = h_{\mu\nu}$ and $\partial \phi = \psi$. One interesting nature of this wave function is that there is only one boundary (final boundary) of the path integral, while usually, the path integral must have two boundaries (initial and final boundaries). Since there is no initial boundary of the wave function, this wave function is also known as the \textit{no-boundary} wave function.

\subsection{Steepest-descent approximation and fuzzy instantons}

In the context of cosmology, it is reasonable to assume the $O(4)$-symmetry as follows:
\begin{eqnarray}
ds_{\mathrm{E}}^{2} = d\tau^{2} + a^{2}(\tau) d\Omega_{3}^{2},
\end{eqnarray}
where $\tau$ is the Euclidean time, $d\Omega_{3}^{2}$ is the three sphere, and $a(\tau)$ is the scale factor. In addition to this symmetry, if we impose the on-shell condition to the metric and the matter field, or we restrict to the \textit{instantons}, we can approximate the wave function based on the steepest-descent approximation:
\begin{eqnarray}
\Psi\left[b,\psi\right] \simeq \sum_{\mathrm{on-shell}} e^{-S_{\mathrm{E}}^{\mathrm{on-shell}}},
\end{eqnarray}
where $b$ and $\psi$ are the boundary values of $a(\tau)$ and $\phi(\tau)$, respectively. Finally, the probability for each instanton is approximately,
\begin{eqnarray}
P\left[b,\psi\right] = \left| \Psi\left[b,\psi\right] \right|^{2} \simeq e^{-2\,\mathrm{Re}\,S_{\mathrm{E}}^{\mathrm{on-shell}}},
\end{eqnarray}
where
\begin{eqnarray}
S_{\mathrm{E}}^{\mathrm{on-shell}} = \mathrm{Re}~S_{\mathrm{E}}^{\mathrm{on-shell}} + i~\mathrm{Im}~S_{\mathrm{E}}^{\mathrm{on-shell}}.
\end{eqnarray}

In general, the on-shell solutions can be complex-valued, in other words, instantons are \textit{fuzzy}, but the boundary values $b$ and $\psi$ must be real-valued \cite{Halliwell:1989dy}. In some sense, this is a kind of boundary condition of the instantons. In addition, the reality at the boundary of the wave function is related to the \textit{classicality} of the solution. At once the solution became classical, along the steepest-descent path, the probability must be slowly varied. If the solution is real-valued, or at least if the real component of each function is dominant than that of the imaginary part, then the probability must be slowly varied compared to the phase part, and hence, the history is sufficiently classical; and at the same time, the history should satisfy the classical equations of motion, e.g., the Hamilton-Jacobi equation. This condition can be summarized as follows:
\begin{eqnarray}
\left| \nabla_{\alpha} \mathrm{Re}~S_{\mathrm{E}} \right| \ll \left| \nabla_{\alpha} \mathrm{Im}~S_{\mathrm{E}} \right|,
\end{eqnarray}
where $\alpha =a, \phi$ denotes the canonical direction \cite{Hartle:2008ng}. In many practical cases, this classicality condition can be easily demonstrated by checking whether the real parts of functions are dominated that of the imaginary parts after the Wick-rotation and after a sufficient Lorentzian time.

\subsection{Scope of this paper}

The natural question is this: then, for which physical situation, the classicality condition can be satisfied? The answer is that \textit{inflation is needed} to satisfy the classicality condition. This is very important: if our universe was created from the Hartle-Hawking wave function, then a small amount of inflation is needed \cite{Hartle:2007gi}. However, still there remain several questions as follows.
\begin{itemize}
\item[1.] Does the Hartle-Hawking wave function prefer sufficient inflation?
\item[2.] Which type of inflation allows classicalization? Slow-roll or even including fast-roll?
\item[3.] Is the Hartle-Hawking wave function a unique choice for quantum cosmology, or can there be further generalization from the Euclidean path integral approach?
\item[4.] Is the Hartle-Hawking wave function compatible with the recent progress of quantum gravity?
\end{itemize} 
In this paper, we review several interesting progress about the Hartle-Hawking wave function and its potential applications. In addition to this, we answer the previous questions and provide some possible future applications and research directions.

\section{Fuzzy instantons with slow-roll inflation}

The first issue is to obtain the classicalized fuzzy instantons based on the slow-roll inflation.

\subsection{Simplest model}

To discuss the generic properties of the slow-roll inflation and fuzzy instantons, we consider the following model \cite{Hartle:2008ng}:
\begin{eqnarray}
S_{\mathrm{E}} = - \int d^{4}x \sqrt{+g} \left[ \frac{1}{16\pi} \left( R - 2\Lambda \right) - \frac{1}{2} \left(\nabla\Phi\right)^{2} - V\left(\Phi\right) \right],
\end{eqnarray}
where $R$ is the Ricci scalar, $\Lambda$ is the cosmological constant, $\Phi$ is a scalar field, and
\begin{eqnarray}
V(\phi) = \frac{1}{2} m^{2} \Phi^{2}
\end{eqnarray}
is the potential. For simplicity, one can define the metric and several variables as follows:
\begin{eqnarray}
ds_{\mathrm{E}}^{2} &=& \frac{3}{\Lambda} \left( d\tau^{2} + a^{2}(\tau) d\Omega_{3}^{2} \right),\\
\phi &\equiv& \sqrt{\frac{4\pi}{3}} \Phi,\\
\mu &\equiv& \sqrt{\frac{3}{\Lambda}} m.
\end{eqnarray}

The equations of motion are as follows:
\begin{eqnarray}
\ddot{a} + a + a \left( 2\dot{\phi}^{2} + \mu^{2} \phi^{2} \right) &=& 0,\\
\ddot{\phi} + 3\frac{\dot{a}}{a}\dot{\phi} - \mu^{2} \phi &=& 0.
\end{eqnarray}
These are two second order differential equations, but we will consider complexified instantons. Hence, each equation has two parts, where one is the real part and the other is the imaginary part. Therefore, basically, there are eight initial conditions (at $\tau = 0$) that decide the solution. However, these eight conditions are not free, but restricted, if we assume the no-boundary condition $a(\tau=0) = 0$. In order to obtain the consistent no-boundary condition, we must assume that
\begin{eqnarray}
a(\tau = 0) &=& 0,\\
\dot{a}(\tau = 0) &=& 1,\\
\dot{\phi}(\tau = 0) &=& 0,
\end{eqnarray}
where these equations already fixed six initial conditions. There are two free parameters $\mathrm{Re}~\phi(\tau = 0)$ and $\mathrm{Im}~\phi(\tau = 0)$, or we present
\begin{eqnarray}
\phi(\tau = 0) = \phi_{0} e^{i\theta},
\end{eqnarray}
where both of $\phi_{0}$ and $\theta$ are real values.

\begin{figure}
\widefigure
\includegraphics[scale=0.3]{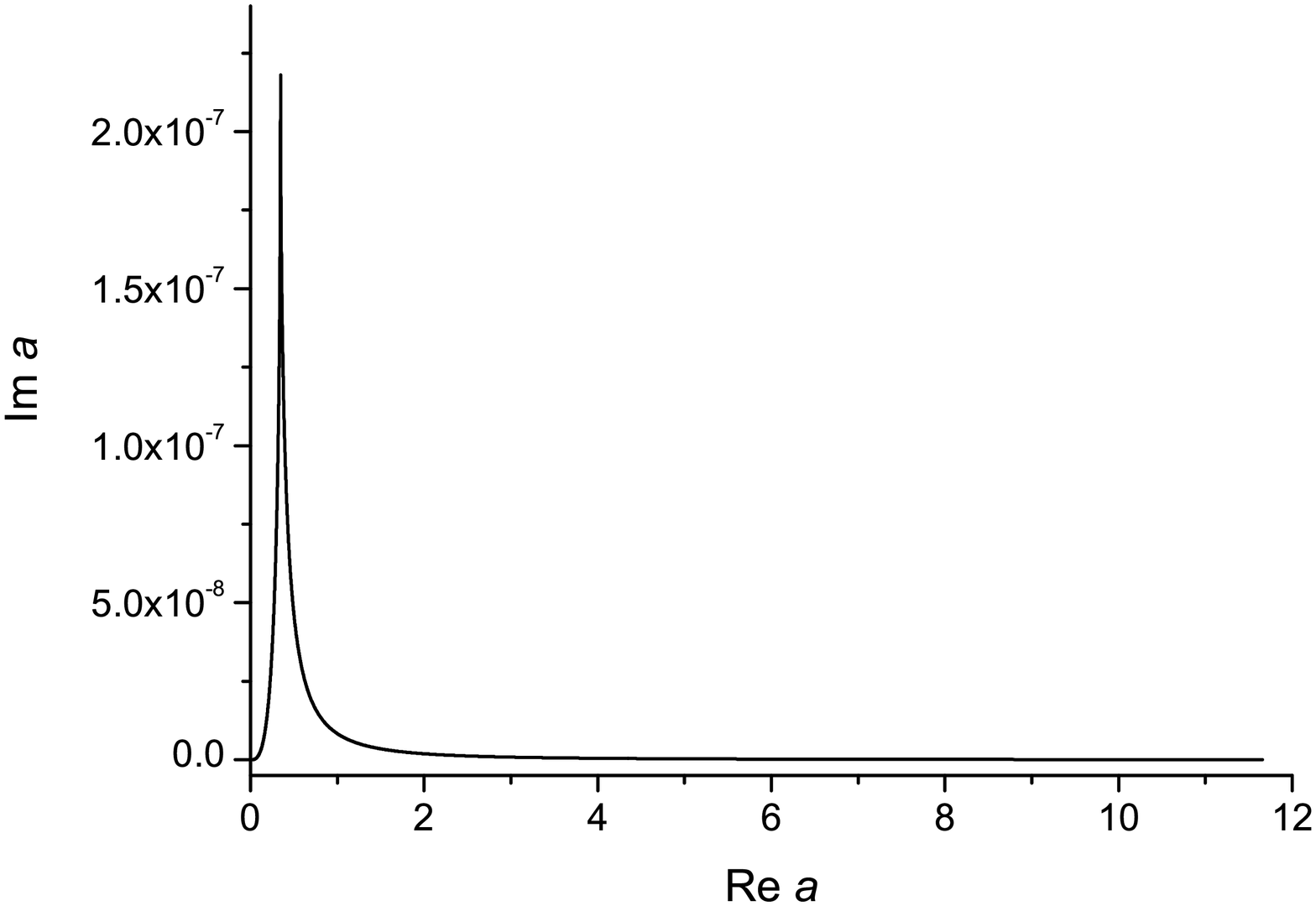}
\includegraphics[scale=0.3]{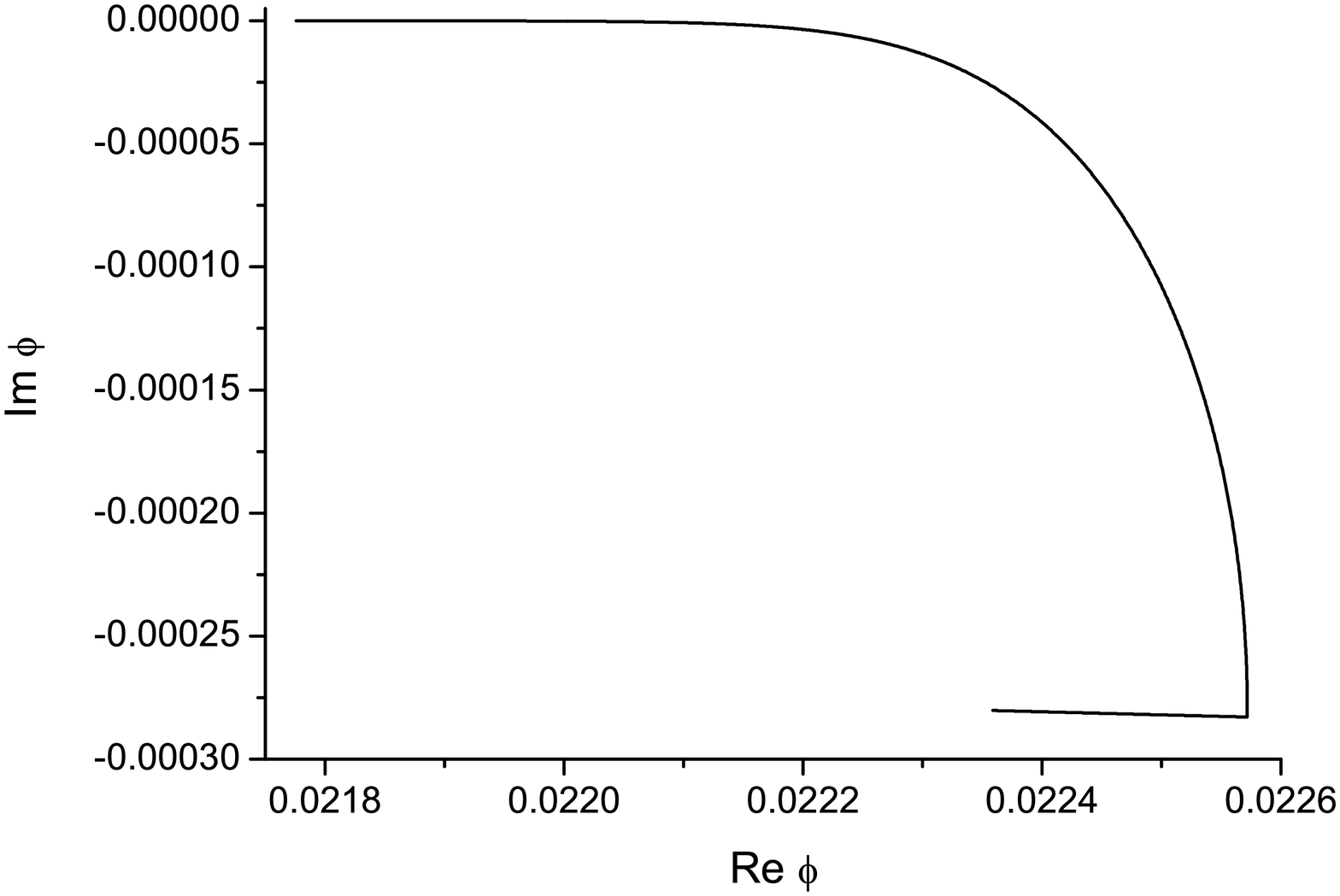}
\caption{\label{fig:rho}An example of a fuzzy instanton solution with $m^{2}/V_{0} = 0.2$ and $m\phi_{0}/\sqrt{V_{0}}=0.02$, where left is the metric $a$ and right is the scalar field $\phi$. Here, the cusp is the turning point from Euclidean to Lorentizan signatures \cite{Hwang:2012mf}.}
\end{figure}

Physically, $\phi_{0}$ corresponds to the initial condition of the inflaton field that decides the structure of each universe. Therefore, we will eventually see the probability distribution as a function of $\phi_{0}$. On the other hand, $\theta$ is just a free parameter. This must be used to satisfy the boundary condition after the Wick-rotation, i.e., for the classicality. If we choose a proper $\theta$, then it may be possible that, after the Wick-rotation $\tau = \tau_{0} + it$, the imaginary parts of both $a$ and $\phi$ approach to zero, while the real parts are dominated (for example, see Fig.~\ref{fig:rho}, \cite{Hwang:2012mf}). Therefore, in other words, $\theta$ is a tuning parameter for the classicality at future infinity.

\end{paracol}
\begin{paracol}{2}
\linenumbers
\switchcolumn

Then, one may ask this: what is the detailed role of $\theta$ \cite{Hwang:2014vba}? In order to see this, let us first assume the slow-roll inflation in a classical background metric, i.e., $\dot{\phi}^{2} \ll 1$, $\mu^{2}\phi^{2} \ll 1$, and $\mathrm{Im}~a \ll \mathrm{Re}~a$. In this case, one can write that
\begin{eqnarray}
\mathrm{Re}~a = C_{a} e^{t} + D_{a} e^{-t} \simeq C_{a} e^{t},
\end{eqnarray}
where $C_{a}$ and $D_{a}$ are integration constants. The equation of motion for the scalar fields are approximately
\begin{eqnarray}
\mathrm{Re}~\ddot{\phi} + 3 \mathrm{Re}~\dot{\phi} + \mu^{2} \mathrm{Re}~\phi &\simeq& 0,\\
\mathrm{Im}~\ddot{\phi} + 3 \mathrm{Im}~\dot{\phi} + \mu^{2} \mathrm{Im}~\phi &\simeq& 0.
\end{eqnarray}
Hence,
\begin{eqnarray}
\mathrm{Re}~\phi &\simeq& C^{\mathrm{Re}}_{\phi} e^{-\frac{3}{2}t + \omega t} + D^{\mathrm{Re}}_{\phi} e^{-\frac{3}{2}t - \omega t},\\
\mathrm{Im}~\phi &\simeq& C^{\mathrm{Im}}_{\phi} e^{-\frac{3}{2}t + \omega t} + D^{\mathrm{Im}}_{\phi} e^{-\frac{3}{2}t - \omega t},
\end{eqnarray}
where $C^{\mathrm{Re}}_{\phi}$, $D^{\mathrm{Re}}_{\phi}$, $C^{\mathrm{Im}}_{\phi}$, and $D^{\mathrm{Im}}_{\phi}$ are constants and
\begin{eqnarray}
\omega^{2} \equiv \left(\frac{3}{2}\right)^{2} - \mu^{2}.
\end{eqnarray}

From these equations, we can easily obtain the following conclusion. If $\mu < 3/2$, then there is a way to choose $\theta$ (hence, finely tune integration constants) and obtain $C^{\mathrm{Im}}_{\phi} \ll 1$. Then,
\begin{eqnarray}
\frac{\mathrm{Im}~\phi}{\mathrm{Re}~\phi} \simeq e^{-2\omega t} \rightarrow 0,
\end{eqnarray}
and hence, after the Wick-rotation, the solution will eventually satisfy the classicality conditions. On the other hand, if $\mu > 3/2$, then for any choice of the initial condition,
\begin{eqnarray}
\frac{\mathrm{Im}~\phi}{\mathrm{Re}~\phi} \simeq  \mathcal{O} \left(1\right).
\end{eqnarray}
Therefore, the imaginary part and the real part of the scalar field are in a similar order.

One may ask what is the consequences if there is no way to classicalize the scalar field. Note that the imaginary part of the scalar field provides the negative kinetic term, and hence, the imaginary part is ghost-like. The energy of the scalar field will turn to the matter content of the universe. If the imaginary part of the scalar field is in the same order as the real part even after the Wick-rotation, then we cannot avoid the instability of the ghost-like imaginary part of the scalar field. This is a catastrophic consequence and physically we cannot allow this possibility. (However, for the possibility to see some restricted contributions from the ghost-like term, see \cite{Chen:2015ria}.)

In a generic scalar field potential $V$, the criterion for the classical universe is $m^{2} / \Lambda < 9/4$, or
\begin{eqnarray}\label{eq:pot}
\left| \frac{V''}{V} \right| < 6\pi.
\end{eqnarray}
If $m^{2}/V < 6\pi$, then $\phi = 0$ (local minimum) can have the classical universe. If $m^{2}/V > 6\pi$, there exists a cutoff $\phi_{\mathrm{cutoff}} > 0$ such that $\phi > \phi_{\mathrm{cutoff}}$ only allows the classical universes \cite{Hartle:2008ng}.

\subsection{Probabilities and the preference of large $e$-foldings}

At once one can construct a classical universe, it is possible to obtain the probability. For the slow-roll potential, the probability of a classical universe with the initial condition $\phi_{0}$ is approximately
\begin{eqnarray}
\log P \simeq \frac{3}{8V(\phi_{0})}.
\end{eqnarray}
Note that this is positive definite. Hence, the probability is exponentially enhanced. The most preferred initial condition is the field value of the smallest potential. This implies that $\phi \simeq \phi_{\mathrm{cutoff}}$ corresponds the most probable initial condition from the no-boundary proposal \cite{Hartle:2007gi}.

However, the problem is that the $e$-foldings of the most probable initial condition are not sufficient in general. For example, if we have the quadratic potential with $\Lambda = 0$, the most preferred $e$-folding number is $\sim 0.62$, while we need more than $\sim 50$ $e$-foldings \cite{Hwang:2012bd}.

There have been several proposals to resolve or understand this problem. First, the simplest opinion is that the no-boundary proposal is simply wrong. For example, if we do not trust the steepest-descent approximation for some theoretical reasons, we may obtain an alternative probability distribution \cite{Feldbrugge:2017kzv}. Or, if we start from a new fundamental wave function, it is possible to obtain a different probability distribution that may prefer large vacuum energy \cite{Vilenkin:1987kf}. However, it is fair to say that there are still several theoretical arguments to support the consistency of the original approaches of Hartle and Hawking \cite{DiazDorronsoro:2017hti}. Keeping this in mind, we may consider several viable possibilities to rescue the Hartle-Hawking wave function as follows \cite{Hwang:2013nja}:
\begin{itemize}
\item[1.]\textit{We need some ad hoc terms to measure the probability}. For example, Hartle, Hawking, and Hertog introduced the volume-weighting factor for the probability measure \cite{Hartle:2007gi}. Then, there is a competition between the volume-weighting sector and the Euclidean probability sector; if the vacuum energy is sufficiently large, then the volume-weighting sector becomes dominated, and eventually, large $e$-foldings are preferred. However, this assumption cannot be justified from the first principles.
\item[2.] \textit{Our universe was started from $V \sim 1$ (Planck scale) vacuum energy}. If this is the case, then there are no serious probability differences between the cutoff and the other field values. However, comparing to observational constraints, this Planck-scale inflation cannot be the primordial inflation of our universe. 
\item[3.] \textit{We need unknown physical degrees of freedom}. For example, if there exists a very long field space or a large number of fields that contribute to inflation \cite{Hwang:2012bd}, then such degeneracy of the field space can compete with the Euclidean probability sector. However, in many cases, this requires too many degrees of freedom; hence, these possibilities may be unnatural in terms of the quantum field theory.
\item[4.] \textit{Some modifications of the gravity theory can explain large $e$-foldings}. For example, massive gravity models \cite{Sasaki:2013aka,Zhang:2014wia} can provide some interesting possibilities. However, this approach goes beyond the regime of Einstein gravity.
\end{itemize}

\subsection{Rescue from the secondary scalar field}

However, probably, the most reasonable rescue is to introduce one additional massive field \cite{Hwang:2014vba}. The main idea is that the classicality condition requires the classicality for \textit{all} fields; not only an inflaton field but also other matter fields. If this does not happen, i.e., if the only inflaton field is classicalized while the second field is not classicalized, then the ghost-like modes of the secondary field cannot be controlled after the Wick-rotation; hence, we must avoid this possibility.

For simplicity, let us consider the following model
\begin{eqnarray}
S_{\mathrm{E}} = - \int d^{4}x \sqrt{+g} \left[ \frac{1}{16\pi} R - \frac{1}{2} \left(\nabla\Phi_{1}\right)^{2} - \frac{1}{2} \left(\nabla\Phi_{2}\right)^{2} - \frac{1}{2} m_{1}^{2} \Phi_{1}^{2} - \frac{1}{2} m_{2}^{2} \Phi_{2}^{2} \right],
\end{eqnarray}
where $m_{1}$, $m_{2}$ are mass parameters of $\Phi_{1}$, $\Phi_{2}$, respectively. Especially, we assume that $m_{1} \ll m_{2}$, and hence, $\Phi_{1}$ is the inflaton field and $\Phi_{2}$ is just an assisting field. Similar to the previous section, one can choose the metric ansatz as follows:
\begin{eqnarray}
ds_{\mathrm{E}}^{2} = \frac{1}{m_{2}^{2}} \left( d\tau^{2} + a^{2}(\tau) d\Omega_{3}^{2} \right).
\end{eqnarray}

Because of the slow-roll condition, along the field direction of $\Phi_{2}$, the variation of $\Phi_{1}$ is negligible. Hence, it is possible to approximate that $(1/2)m_{1}\Phi_{1}^{2} \simeq V_{0}$ is a constant. Therefore, the classicality condition of the potential (Eq.~(\ref{eq:pot})) is
\begin{eqnarray}
\frac{m_{2}^{2}}{V_{0}} \simeq \frac{m_{2}^{2}}{\frac{1}{2} m_{1}^{2} \Phi_{1}^{2}} \leq 6\pi.
\end{eqnarray}
The numerical investigation is also consistent to this expectation (Fig.~\ref{fig:cutoffs_new}, \cite{Hwang:2014vba}). The shadowed box region becomes narrower and narrower as $m_{1}/m_{2}$ decreases. Hence, in the $m_{1}/m_{2} \ll 1$ limit, if $\Phi_{2} \simeq 0$, the genuine cutoff of the $\Phi_{1}$ direction will satisfy $\Phi_{1} \gg 1$.

\end{paracol}
\begin{figure}
\widefigure
\includegraphics[scale=0.4]{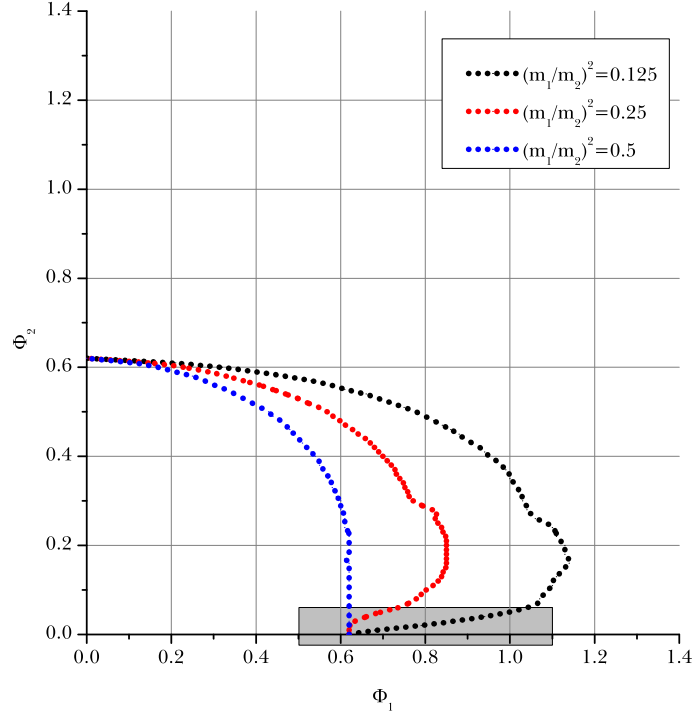}
\caption{\label{fig:cutoffs_new}Numerical calculations of the cutoffs for $(m_{1}/m_{2})^{2} = 0.125, 0.25$, and $0.5$ \cite{Hwang:2014vba}.}
\end{figure}  
\begin{paracol}{2}
\linenumbers
\switchcolumn

Now we need to ask what is the most probable initial condition over the field space $(\Phi_{1},\Phi_{2})$. Of course, the smallest potential energy is the most preferred initial condition. Since we assumed $m_{1} \ll m_{2}$, the potential very sensitively varies along the $\Phi_{2}$ direction. Hence, the initial conditions with $\Phi_{2} \simeq 0$ must be the most preferred direction. Now, if we assume $\Phi_{2} \simeq 0$, then the most probable initial condition of the $\Phi_{1}$ direction is $\Phi_{1,\mathrm{cutoff}}$. However, in order to classicalize the $\Phi_{2}$ field, the following condition must be satisfied:
\begin{eqnarray}
\frac{m_{2}^{2}}{3\pi m_{1}^{2}} \leq \Phi_{1,\mathrm{cutoff}}^{2},
\end{eqnarray}
where the detailed constants in the left-hand-side is not very important. If there is a mass hierarchy $m_{1} \ll m_{2}$, then the cutoff of $\Phi_{1}$ can be sufficiently large, while this means that the initial condition of the inflaton field must have large $e$-foldings $\mathcal{N}$:
\begin{eqnarray}
\mathcal{N} \simeq \mathcal{O}\left(1\right) \times \frac{m_{2}^{2}}{m_{1}^{2}},
\end{eqnarray}
where it is quite easy to make $\mathcal{N}$ greater than $50$.

Of course, this idea is not limited to the quadratic potential models; it might be interesting to apply this idea for realistic inflation models as well as models with various interactions between fields.

\section{Fuzzy instantons with fast-roll potential}

The second issue is to obtain the classicalized fuzzy instantons even if the slow-roll condition is not guaranteed. Indeed, this issue must be highlighted in the recent discussion in string theory.

\subsection{Landscape vs. Swampland}

To understand the cosmological constant problem as well as several fine-tuning issues of the universe, the \textit{cosmic landscape} was a very fancy hypothesis \cite{Susskind:2003kw}. String theory allows various (almost all possible) constants of nature, including the cosmological constant and detailed shapes of the inflaton potential, where this is called by the cosmic landscape. These possible parameter spaces are physically realized via eternal inflation and quantum tunneling of bubble universes. Eventually, in somewhere of the \textit{multiverse}, any fine-tuned parameters can be realized.

Although there are several criticisms about this approach, the most serious criticism was suggested from the string community \cite{Obied:2018sgi}. According to the authors, the landscape, where the string theory allowed, is indeed a very restricted region among the possible parameter spaces. For example, it was conjectured that the inflaton potential must be restricted by
\begin{eqnarray}
\left| \frac{V''}{V} \right| &>& \mathcal{O}(1),\\
\left| \frac{V'}{V} \right| &>& \mathcal{O}(1),
\end{eqnarray}
where this is known to be the \textit{swapland criteria}. Of course, there are several subtle issues. First, there is no fundamental proof about them. Hence, the order one constant is subtle to define. Perhaps, the slow-roll inflation can be marginally allowed \cite{Brahma:2020cpy}, but there may be no fundamental cosmological constant if we seriously accept the swampland criteria.

\end{paracol}
\begin{figure}
\widefigure
\includegraphics[scale=1]{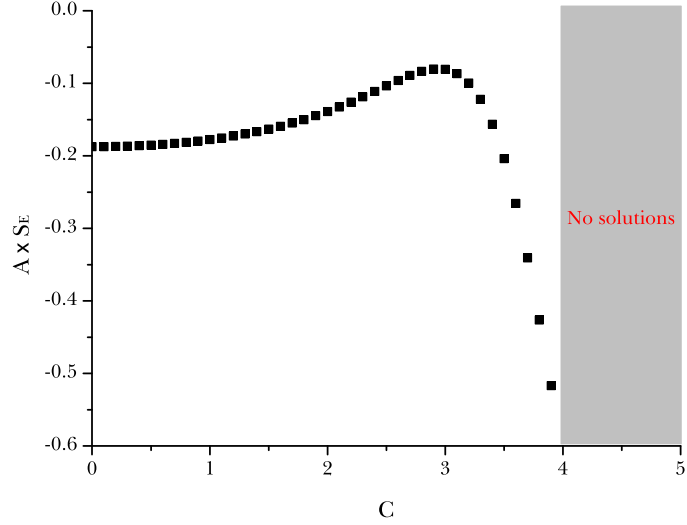}
\caption{\label{fig:exponential}Euclidean action for the exponential potential. If $C>4$, then classicalized solutions are not allowed \cite{Hwang:2012zj}.}
\end{figure}  
\begin{paracol}{2}
\linenumbers
\switchcolumn

In this paper, we do not agree or disagree about the details of the swampland criteria. However, this is already harmful to the Hartle-Hawking proposal; there is a tension to Eq.~(\ref{eq:pot}). We may see more details for a run-away quintessence model which is very typical from string-inspired models \cite{Hwang:2012zj}:
\begin{eqnarray}
V(\phi) = A e^{-C \phi}.
\end{eqnarray}
For each point near $\phi_{0}$, we can approximate the potential such as
\begin{eqnarray}
V(\phi) \simeq \frac{1}{2} A C^{2} e^{-C \phi_{0}} \left( \phi - \phi_{0} - \frac{1}{C}\right)^{2} + \frac{A e^{-C\phi_{0}}}{2}.
\end{eqnarray}
Therefore, it is not surprising that $\phi = \phi_{0}$ has a classical history only if
\begin{eqnarray}
C^{2} \lesssim 3\pi.
\end{eqnarray}
From the numerical computations, we can confirm that $C \lesssim 4$ is the condition for the existence of the classical solutions (Fig.~\ref{fig:exponential}, \cite{Hwang:2012zj}). On the other hand, if $C > 4$ happens which is very natural from the string-inspired models, there are no classicalized instantons along the run-away direction. Hence, such a quintessence model is not compatible with the Hartle-Hawking wave function.

Then, is there any way to rescue the Hartle-Hawking wave function even in the context of the swampland criteria?

\subsection{Rescue from Hwang-Sahlmann-Yeom instantons}

Even though the swampland criteria do not like the local minimum of the potential, these do not exclude the unstable local maximum of the potential. We describe the hill-top potential near the hill-top ($\phi = 0$)
\begin{eqnarray}
V(\phi) = V_{0} \left( 1 - \frac{1}{2} \mu^{2} \phi^{2} \right).
\end{eqnarray}
If $\mu \ll 1$, then the slow-roll condition is satisfied, and the usual fuzzy instanton can exist. On the other hand, if $\mu \gg 1$, which is more natural from string-inspired models, then the slow-roll condition is no more satisfied. So, near the hilltop, during the Euclidean time, the initial field values rapidly emerge to the local maximum of the potential. By choosing proper initial conditions, one could obtain the fuzzy instantons near the fast-rolling hill-top potentials \cite{Hwang:2011mp} (for example, see Fig.~\ref{fig:toy_HSY}). We name these solutions Hwang-Shalmann-Yeom (HSY) instantons to contrast to the slow-roll fuzzy instantons of Hartle-Hawking-Hertog (HHH).

\end{paracol}
\begin{figure}
\widefigure
\includegraphics[scale=0.3]{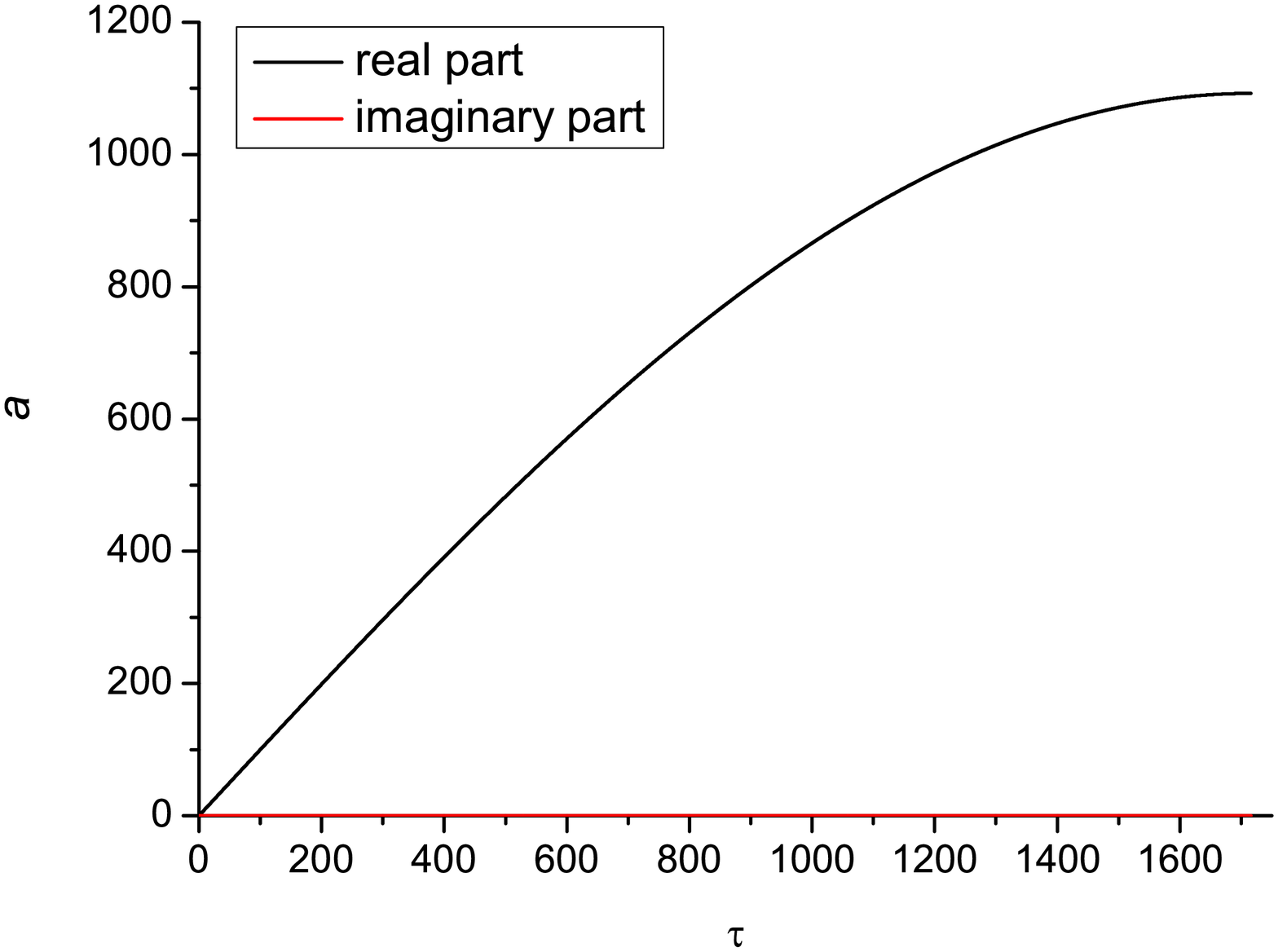}
\includegraphics[scale=0.3]{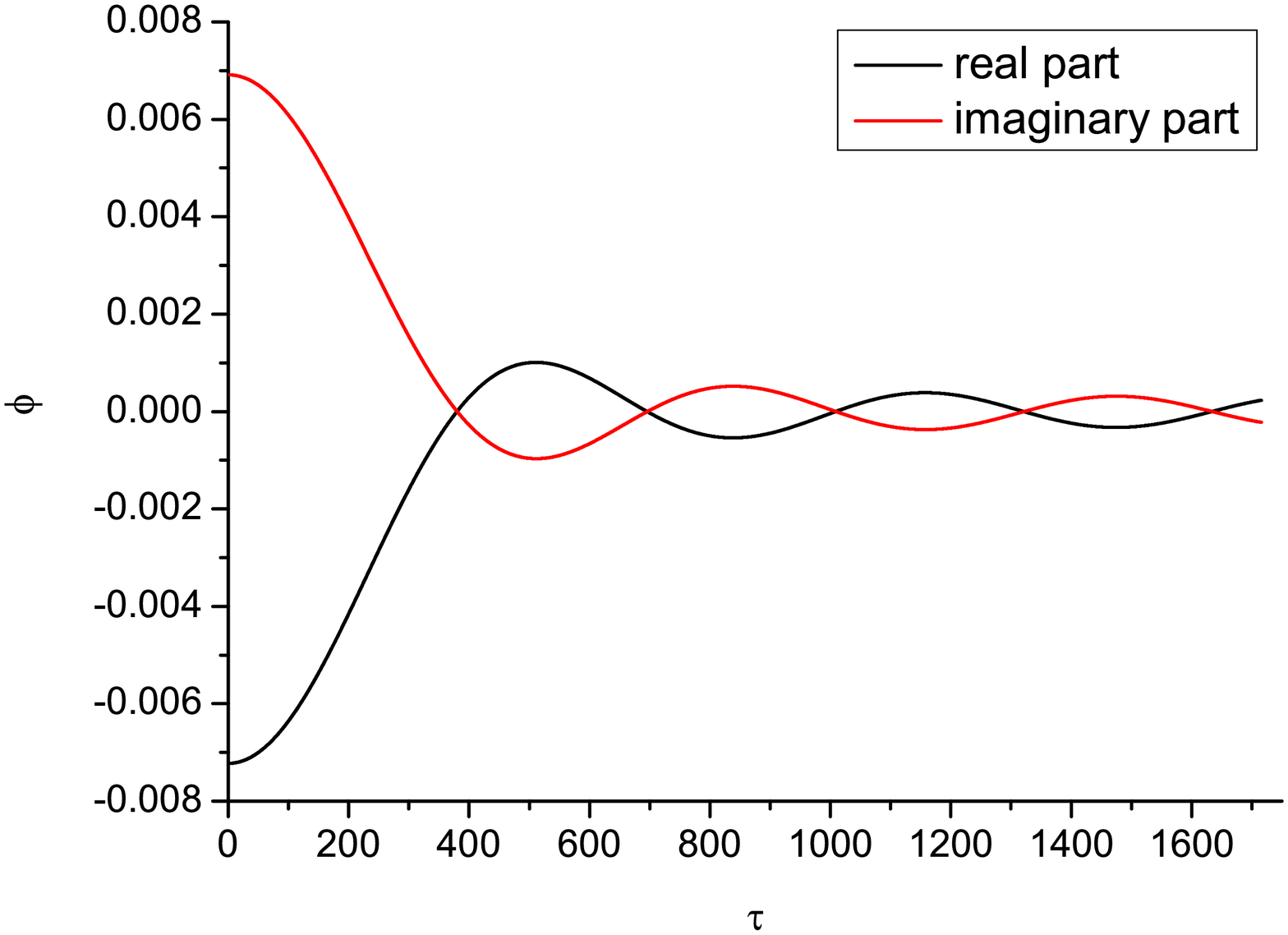}
\includegraphics[scale=0.3]{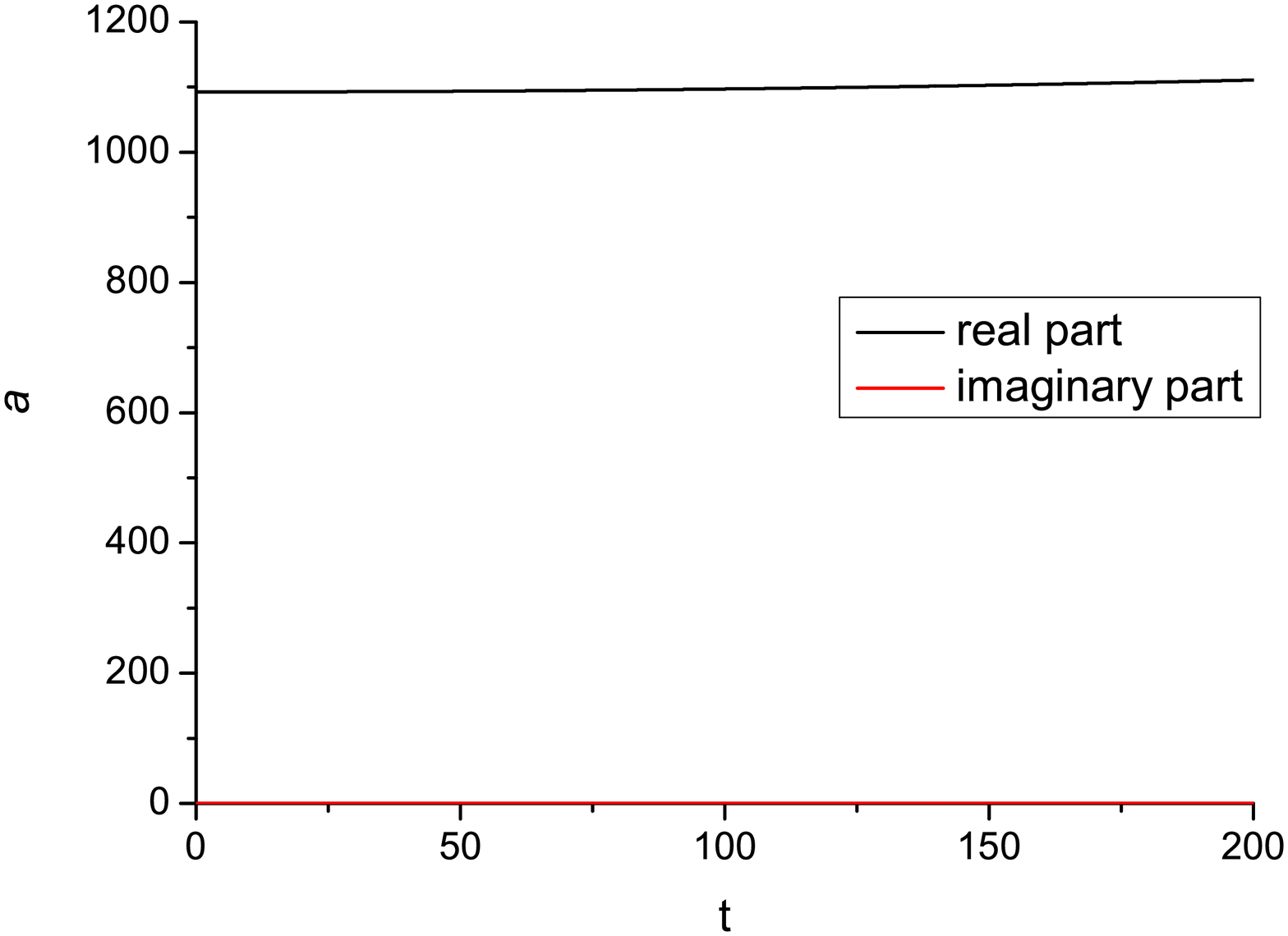}
\includegraphics[scale=0.3]{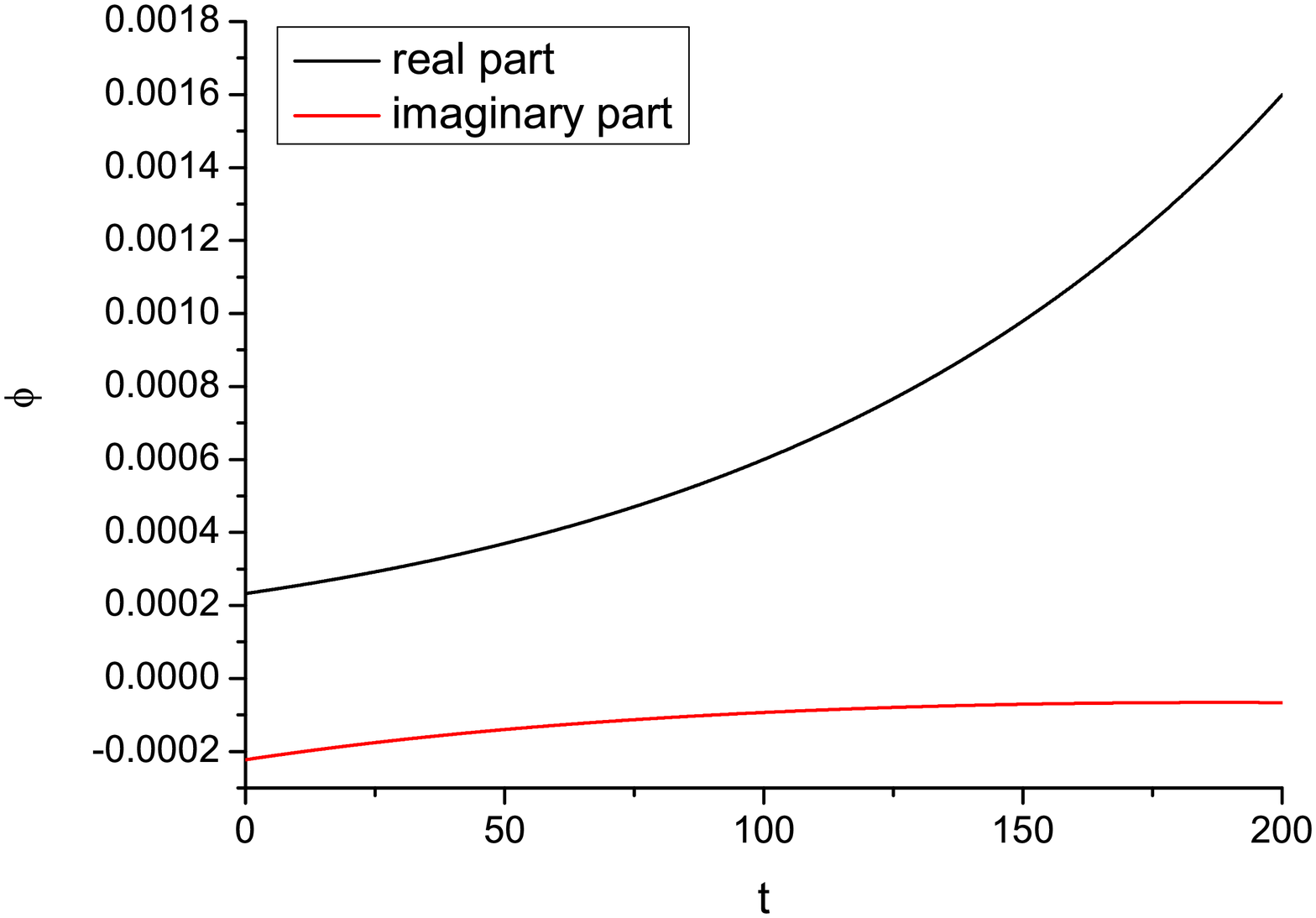}
\caption{\label{fig:toy_HSY}A fuzzy instanton solution ($\phi_{0}=0.01$) of a toy potential $V = V_{0} - (1/2) m^{2} \phi^{2} + (1/24) \lambda \phi^{4}$, where $V_{0}=10^{-7}$, $m^{2}=10^{-4}$, and $\lambda=2 \times 10^{-2}$. Upper figures denote along the Euclidean time, while lower figures denote along the Lorentzian time. During the Euclidean time, the field rapidly oscillates along the hill-top ($\phi = 0$ in this potential, upper right). After the Wick-rotation, the field roles left or right direction (lower right).}
\end{figure}  
\begin{paracol}{2}
\linenumbers
\switchcolumn

The physical difference comes from the probability (Fig.~\ref{fig:transition2}, \cite{Hwang:2012zj}). If $\mu \ll 1$ case, then the probability is approximately
\begin{eqnarray}
\log P_{\mathrm{HHH}} \simeq \frac{3}{8V_{0} \left( 1 - \frac{1}{2}\mu^{2} \phi_{0}^{2} \right)}.
\end{eqnarray}
On the other hand, if $\mu \gg 1$, then the field quickly approaches the local maximum of the potential, and hence, the initial condition dependence is negligible:
\begin{eqnarray}
\log P_{\mathrm{HSY}} \simeq \frac{3}{8V_{0}}.
\end{eqnarray}
Therefore, at once there is a hilltop with $\mu \gg 1$, the probability for left-rolling and right-rolling are almost the same.

\end{paracol}
\begin{figure}
\widefigure
\includegraphics[scale=1]{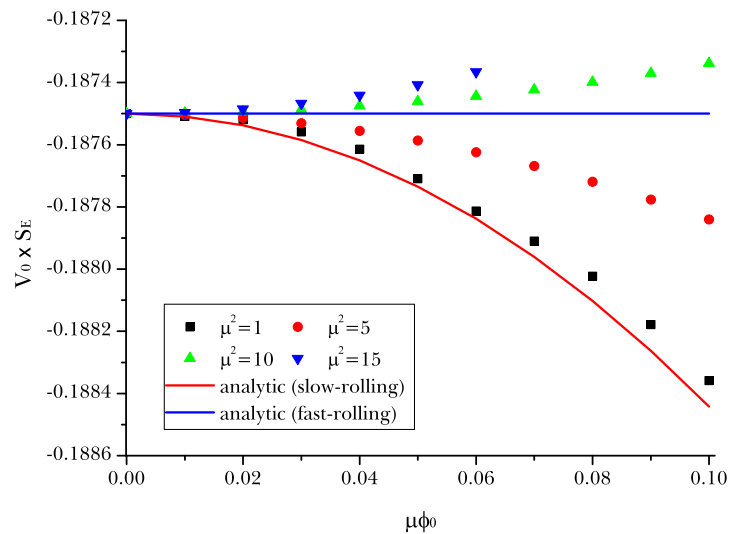}
\caption{\label{fig:transition2}Euclidean action $V_{0} S_{\mathrm{E}}$ as a function of $\mu\phi_{0}$. The red curve is $\log P_{\mathrm{HHH}}$, while the blue line is $\log P_{\mathrm{HSY}}$ \cite{Hwang:2012zj}.}
\end{figure}  
\begin{paracol}{2}
\linenumbers
\switchcolumn

This HSY instanton can rescue the Hartle-Hawking wave function even in the context of the swampland because the classicalized universes can be created near the unstable and sharp hill-top of the field space.

\subsection{Cosmological applications}

At once there exist fuzzy instantons in the context of the cosmic landscape and the swampland criteria, there can be various cosmological applications using HSY instantons \cite{Hwang:2012zj}.
\begin{itemize}
\item[1.] About the moduli or the dilaton stabilization issue, there is a probability competition between the stable direction and the unstable direction. Perhaps, the only possible starting point from the no-boundary wave function is the hilltop of the potential; according to the HSY instantons, there is no preference between the left-rolling and the right-rolling. Hence, with the reasonable probability, the moduli or dilaton stabilization can be explained from quantum cosmology \cite{Hwang:2011mp}.
\item[2.] The universe starts not from the local minimum but the local maximum. The cosmological constant is about the local \textit{minimum}, but the probability of the HSY instanton is about the local \textit{maximum}. Therefore, even though the cosmological constant varies from anti-de Sitter to de Sitter, there might be no singular changes of the \textit{a prior} probability, because there is no singular change of the local maximum \cite{Hwang:2012zj}.
\item[3.] HSY instantons can rescue the Hartle-Hawking wave function even in the context of the swampland because the classicalized universes can be created near the unstable and sharp hilltop of the field space.
\end{itemize}

To embed a consistent inflation model with the swampland criteria as well as the trans-Planckian censorship conjecture, if we only consider the single-field inflation model, then the no-boundary wave function is not compatible with them \cite{Brahma:2020cpy}. On the other hand, if we include one more field, there might be a possibility to rescue the no-boundary proposal. Or, if the universe was started from the hilltop of the very sharp potential, then it can be explained from the no-boundary wave function; however, the smooth connection to the successful inflation model must be explained.

\section{Extensions}

In previous sections, we have investigated the no-boundary proposal in a single scalar field model with Einstein gravity. However, in principle, there can be further extensions. For example, the following directions are possible.
\begin{itemize}
\item[1.] The Euclidean path integral does \textit{not} necessarily imply the no-boundary proposal. The no-boundary proposal is a specific choice of the Euclidean path integral. In more generic cases, there can be two boundaries (initial and final boundaries); however, due to the ambiguity of time in quantum gravity, one may further interpret that two universes are created from nothing. These solutions are known to be Euclidean wormholes (Fig.~\ref{fig:bigsol}, \cite{Chen:2016ask}).
\item[2.] The theory can be extended or embedded by \textit{quantum gravitational models}. For example, string-inspired models can introduce some additional terms, e.g., the Gauss-Bonnet term with a dilaton coupling \cite{Chew:2020lkj}. Or, the loop quantum cosmological models can provide the big bounce near the putative singularity \cite{Brahma:2018elv}. These corrections can introduce a new type of solution.
\end{itemize}

\subsection{Fuzzy Euclidean wormholes}

As a simple extension, we consider fuzzy Euclidean wormholes in Einstein gravity \cite{Chen:2016ask}. Indeed, in terms of instantons, the Euclidean wormholes are more natural than the compact instantons. The intuitive reason is as follows \cite{Chen:2017qeh}.

\end{paracol}
\begin{figure}
\widefigure
\includegraphics[scale=0.3]{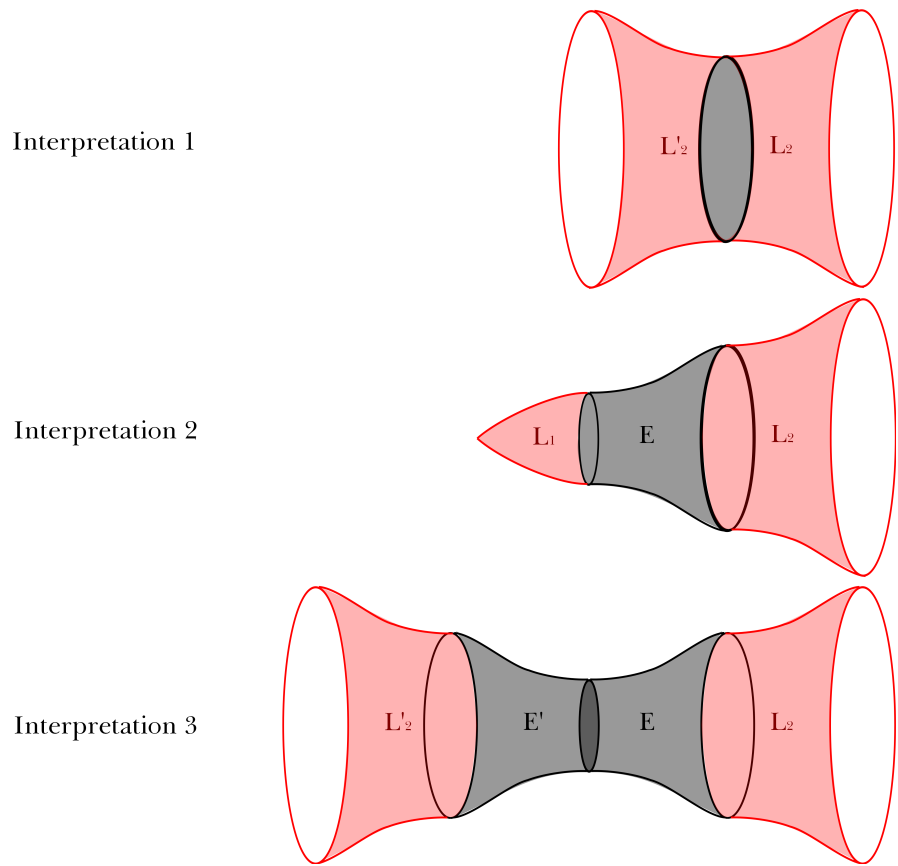}
\caption{\label{fig:bigsol}Possible interpretations of Euclidean wormholes \cite{Chen:2016ask}. Upper: A collapsing universe is bounced (Interpretation 1). Middle: One can interpret that a small universe tunnels to a large universe or one contracting and one expanding universes are created (Interpretation 2). Lower: One can also interpret that two entangled universes are created or a contracting universe is bounced to an expanding universe (Interpretation 3).}
\end{figure}  
\begin{paracol}{2}
\linenumbers
\switchcolumn

Let us first consider a free scalar field model with $O(4)$-symmetric metric ansatz. Then, the following is the generic solution of the scalar field in the Lorentzian signatures:
\begin{eqnarray}
\frac{d\phi}{dt} = \frac{\mathcal{A}}{a^{3}}.
\end{eqnarray}
Of course, because of the classicality, $\mathcal{A}$ is a real-valued number. If we Wick-rotate this solution to the Euclidean time, then we obtain
\begin{eqnarray}
\frac{d\phi}{d\tau} = -i \frac{\mathcal{A}}{a^{3}}.
\end{eqnarray}
Therefore, if the velocity of the scalar field is non-vanishing in the Lorentzian signatures, and if the solution is classical in the Lorentzian signatures, it is necessary that the velocity of the scalar field must be pure imaginary in the Euclidean signatures. However, the pure imaginary scalar field in Euclidean signatures are perfectly acceptable in the Euclidean path integral formalism. Then, the corresponding Euclidean metric satisfies
\begin{eqnarray}
\dot{a}^{2} = 1 - \frac{a^{2}}{\ell^{2}} + \frac{a_{0}^{4}}{a^{4}},
\end{eqnarray}
where $\ell \equiv \sqrt{3/\Lambda}$ and $a_{0}^{4} = 4\pi \mathcal{A}^{2}/3$. 

If $a_{0} \ll \ell$, then there are two zeroes of $\dot{a}$. Hence, the Euclidean solution has two turning points, say $a_{\mathrm{max}}$ and $a_{\mathrm{min}}$. If we consider a solution that covers $a_{\mathrm{max}}$ to $a_{\mathrm{min}}$ to $a_{\mathrm{max}}$, this becomes the Euclidean wormhole solution, where there are two boundaries from the solution and the Wick-rotation can be applied for these two boundaries. The compact instantons are available only if $\mathcal{A} = 0$, or when the velocity of the scalar field is zero. On the other hand, the non-compact instantons occur in more general situations when the Lorentzian solutions have non-trivial velocities.

What we have considered is the case when the scalar field is free. The nature question is, what happens if we generalize to a specific inflation model. In order to do this, we can introduce the ansatz of the initial condition of the Euclidean wormholes as follows \cite{Chen:2017qeh}:
\begin{eqnarray}
\mathrm{Re}~a(0) &=& a_{\mathrm{min}} \cosh \eta,\\
\mathrm{Im}~a(0) &=& a_{\mathrm{min}} \sinh \eta,\\
\mathrm{Re}~\dot{a}(0) &=& \sqrt{\frac{4\pi}{3}} \frac{\mathcal{B}}{a_{\mathrm{min}}^{2}} \sqrt{\sinh\zeta \cosh\zeta},\\
\mathrm{Im}~\dot{a}(0) &=& \sqrt{\frac{4\pi}{3}} \frac{\mathcal{B}}{a_{\mathrm{min}}^{2}} \sqrt{\sinh\zeta \cosh\zeta},\\
\mathrm{Re}~\phi(0) &=& \phi_{0} \cos \theta,\\
\mathrm{Im}~\phi(0) &=& \phi_{0} \sin \theta,\\
\mathrm{Re}~\dot{\phi}(0) &=& \frac{\mathcal{B}}{a_{\mathrm{min}}^{3}} \sinh\zeta,\\
\mathrm{Im}~\dot{\phi}(0) &=& \frac{\mathcal{B}}{a_{\mathrm{min}}^{3}} \cosh\zeta,
\end{eqnarray}
where $a_{\mathrm{min}}$, $\mathcal{B}$, $\phi_{0}$, $\eta$, $\zeta$, and $\theta$ are free parameters. However, these free parameters are not entirely free, but should satisfy the real part and the imaginary part equations of the Hamiltonian constraint:
\begin{eqnarray}
0 &=& 1 + \frac{8\pi}{3} a_{\mathrm{min}}^{2} \left( - V_{\mathrm{r}} + \sinh 2\eta V_{\mathrm{i}} \right) - \frac{4\pi \mathcal{B}^{2}}{3a_{\mathrm{min}}^{4}} \left( 1 + \sinh 2\zeta \sinh 2\eta \right),\\
0 &=& a_{\mathrm{min}}^{6} + \frac{\mathcal{B}^{2} \sinh 2\eta}{2\left(V_{\mathrm{r}} \sinh 2\eta  + V_{\mathrm{i}}\right)},
\end{eqnarray}
where $V_{\mathrm{r}}$ and $V_{\mathrm{i}}$ are the real and the imaginary part of $V(\phi)$ at $\tau = 0$. Because of these two equations, indeed two parameters among five free parameters are decide; say, $a_{\mathrm{min}}$ and $\eta$.

Now the remained free parameters are $\mathcal{B}$, $\phi_{0}$, $\zeta$, and $\theta$. However, $\mathcal{B}$ determines the amplitude of the imaginary part of the scalar field, and hence, intuitively decides the size of the wormhole throat. $\zeta$ is the parameter that decides the symmetry between the left side and the right side of the wormhole. These two parameters are nothing but the parameters that decide the shape of the wormhole. $\phi_{0}$ corresponds to the initial field value of the solution. Therefore, the only tuning parameter that can be used for the classicality condition is $\theta$.

This situation is the same as the compact instanton case, but there is a serious problem. In the compact instanton case, there is only one boundary (future boundary), and hence, the classicality must be imposed only one boundary. Using $\theta$, we could make one boundary classical. However, in the Euclidean wormhole case, there are two boundaries that we need to classicalize. Hence, in general, it is impossible to classicalize two boundaries at the same time.

However, there can be some exceptional cases. For example, let us consider the following potential
\begin{eqnarray}\label{eq:Vtanh}
V(\phi) = \frac{3}{8\pi\ell^{2}} \left( 1 + A \tanh^{2} \frac{\phi}{\alpha} \right),
\end{eqnarray}
where $\ell$, $A$, and $\alpha$ are free parameters (See Fig.~\ref{fig:FIG1}, \cite{Chen:2017qeh}). This model provides a flat hilltop \cite{Chen:2019cmw}, which is consistent to the Starobinsky model \cite{Starobinsky:1980te} which is preferred by the recent Planck data analysis \cite{Akrami:2018odb}. In this model, the scalar field of the end of the wormhole just rolls down to the local minimum, and hence the primordial inflation is naturally terminated. By tuning $\theta$, we classicalize this end. On the other hand, it is not possible to tune the other end, but if the scalar field rolls up to the hilltop, then it is possible to automatically stop as long as the hilltop is sufficiently flat. If the field stops at the hilltop, the field is not classicalized, but the metric is classicalized because there is no contribution of the kinetic terms of the scalar field. In addition, due to the flat potential, locally, there is a shift symmetry; by shifting the field along the complex direction, one can classicalize the field value at the hilltop. In any case, the point is that we can classicalize one end definitely, and the other end is a little bit subtle, but the Euclidean action does not vary after the Wick-rotation at the hilltop. Since we are observing only one universe, we do not need to care about the very details of the other end of the wormhole, as long as the real part of the Euclidean action is bounded well.

\end{paracol}
\begin{figure}
\widefigure
\includegraphics[scale=0.25]{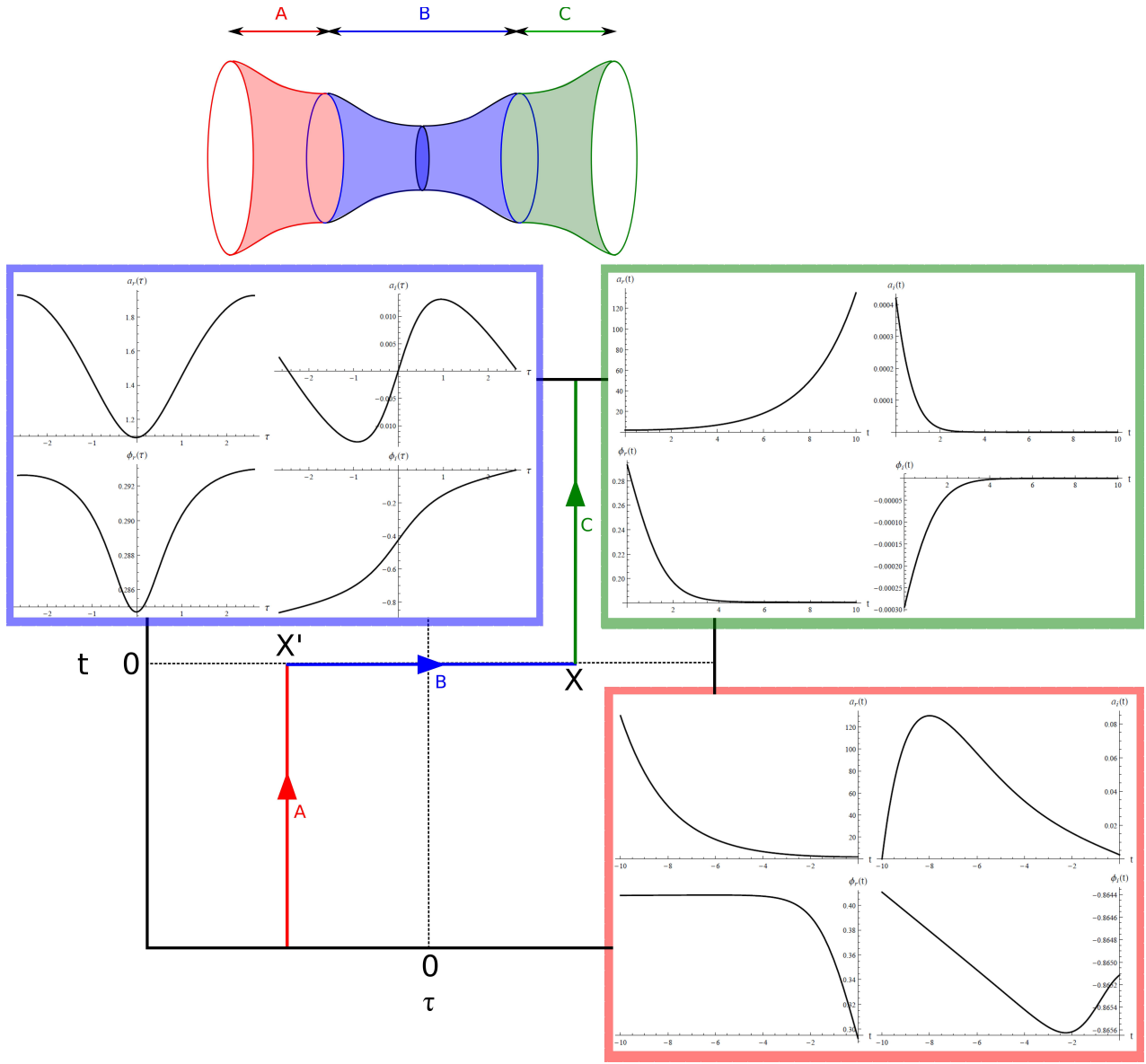}
\caption{\label{fig:FIG1}Complex time contour and numerical solution of $\mathrm{Re}~a$, $\mathrm{Im}~a$, $\mathrm{Re}~\phi$, and $\mathrm{Im}~\phi$ for Eq.~(\ref{eq:Vtanh}). The upper figure is the physical interpretation about the wormhole, where Part A (red) and C (green) are Lorentzian and Part B (blue) is Euclidean \cite{Chen:2017qeh}.}
\end{figure}  
\begin{paracol}{2}
\linenumbers
\switchcolumn

Regarding this mechanism, we have three important comments:
\begin{itemize}
\item[1.] This mechanism cannot be applied for the convex inflaton potential (e.g., quadratic potential). Therefore, the Euclidean wormhole selects the concave inflaton potential \cite{Chen:2017qeh}.
\item[2.] For a given vacuum energy scale $V_{0}$ or $\ell$, the probability of the Euclidean wormhole is larger than that of the compact instanton \cite{Chen:2016ask}, because the maximum probability of the Euclidean wormhole is
\begin{eqnarray}
\log P \simeq \pi \ell^{2} \left( 1 + 0.16 \left(\frac{a_{0}}{\ell}\right)^{5/2} \right),
\end{eqnarray}
where that of the compact instanton is $\log P = \pi \ell^{2}$.
\item[3.] For a given concave potential, there can be a competition between the compact instantons and Euclidean wormholes. The largest probability of the compact instanton occurs near the cutoff, where the probability near the cutoff is in general larger than that of the Euclidean wormholes that only appear near the hilltop. On the other hand, if we assume the mechanism that enhances large $e$-foldings (e.g., introducing a massive field direction), then the Euclidean wormholes are more preferred than that of the compact instantons \cite{Chen:2019cmw}. Therefore, as long as we assume that our universe experienced more than $50$ $e$-foldings, the \textit{Euclidean wormhole} of the \textit{concave} potential is preferred than the compact instantons of the convex or concave potentials.
\end{itemize}

In conclusion, it is interesting that the Euclidean wormholes can answer the question: \textit{why our universe started from a concave part of the potential rather than a convex part}? However, there is one remarkable warning point. In all computations of Euclidean wormholes, we implicitly assumed the UV-completion of the inflaton potentials. In general, the Euclidean wormhole requires the flat direction of the potential. However, in the context of the swampland criteria, this might be an unjustifiable assumption. Finding a sufficiently flat field direction within the UV-completed model is an interesting future research topic.

\subsection{Euclidean wormholes in Gauss-Bonnet-dilaton gravity}

If we regard the string theory as the UV-completion of quantum gravity, it is reasonable to include higher-order corrections of the string-inspired models and see its physical applications. The most famous model is known by the Gauss-Bonnet-dilaton gravity:
\begin{eqnarray}
S = \int d^{4}x \sqrt{-g} \left( \frac{R}{16\pi} - \frac{1}{2} \left( \nabla \phi \right)^{2} - V(\phi) + \frac{1}{2}\xi(\phi) R_{\mathrm{GB}}^{2} \right),
\end{eqnarray}
where
\begin{eqnarray}
R_{\mathrm{GB}}^{2} = R_{\mu\nu\rho\sigma}R^{\mu\nu\rho\sigma} - 4 R_{\mu\nu}R^{\mu\nu} + R^{2}
\end{eqnarray}
is the Gauss-Bonnet term, $\phi$ is the dilaton field, and
\begin{eqnarray}
\xi(\phi) = \lambda e^{-c\phi}
\end{eqnarray}
is the coupling function of the dilaton field. Here, $\lambda$ and $c$ are model-dependent parameters.

Due to the corrections of the Gauss-Bonnet-dilaton term, even though the null energy condition is satisfied, effectively the null curvature condition is violated; or, equivalently, if we regard the Gauss-Bonnet-dilaton term as an effective matter contribution, the null energy condition can be effectively violated. Because of this, it is not surprising that a Lorentzian or Euclidean wormhole solution can exist \cite{Kanti:2011jz}.

In order to obtain a Euclidean wormhole solution, we consider the following initial condition at $\tau = 0$ \cite{Tumurtushaa:2018agq}:
\begin{eqnarray}
a(0) &=& \sqrt{\frac{3}{4\pi \left(2V_{0} - \dot{\phi}^{2}(0) \right)}},\\
\dot{a}(0) &=& 0,\\
\phi(0) &=& \phi_{0},\\
\dot{\phi}(0) &=& 0,
\end{eqnarray}
where $a(0)$ is given from the Hamiltonian constraint equation. By tuning $\phi_{0}$, we must satisfy the boundary condition at $\tau = \tau_{\mathrm{end}}$:
\begin{eqnarray}
a(\tau_{\mathrm{end}}) &=& 0,\\
\dot{a}(\tau_{\mathrm{end}}) &=& -1,
\end{eqnarray}
in order to present the regular end.

In general, this solution penetrates over a sharp potential barrier. Also, since the volume is greater than that of the usual compact instanton, the probability is higher than that of the pure de Sitter instanton. Therefore, at once there exists a string-inspired term, even though there exists a potential barrier, it can be used to create a universe with a higher probability.

If we Wick-rotate at $\tau = 0$, then we can apply this solution for the quantum cosmology \cite{Chew:2020lkj}. However, if we further extend the Euclidean time to $\tau < 0$, and Wick-rotate the solution along the anisotropic direction, then this can explain an (expanding) Lorentzian wormhole from quantum tunneling \cite{Tumurtushaa:2018agq}. The investigation of quantum tunneling of Lorentzian wormhole is also an interesting issue and we need further investigations.

\subsection{Hartle-Hawking wave function with loop quantum cosmology}

In loop quantum gravity, we consider the generic quantum state that satisfies the (quantum) Hamiltonian constraint equation as well as the (quantum) momentum constraint equations. The generic states that satisfy the momentum constraint equations should follow the loop representations. Thanks to the loop representation, there must be a correction to the Hamiltonian constraint in the classical level \cite{Bojowald:2018xxu}. As we include these corrections, we can investigate the effects of quantum gravity.

\end{paracol}
\begin{figure}
\widefigure
\includegraphics[scale=0.9]{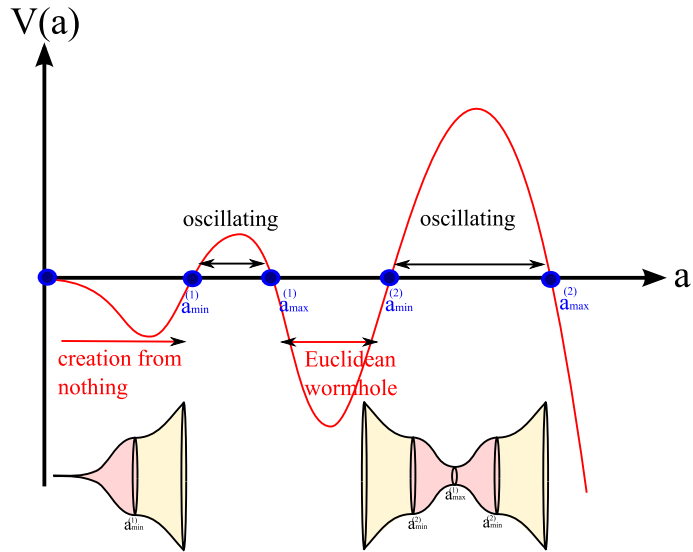}
\caption{\label{fig:pots}A conceptual interpretation of $\mathcal{V}(a)$ \cite{Brahma:2018elv}.}
\end{figure}  
\begin{paracol}{2}
\linenumbers
\switchcolumn

In general, it is believed that, in a cosmological context, the beginning of the universe can be explained by the big bounce. The Lorentzian dynamics of the scale factor satisfies the equation $\dot{a}^{2} + \mathcal{V}(a) = 0$, where $\mathcal{V}(a)$ has a zero at a minimum value $a_{\mathrm{min}}$; this corresponds the bouncing point of the universe \cite{Brahma:2018elv}. However, still there remains a conceptual question: as $a$ goes to $a_{\mathrm{min}}$, the universe approaches the deep quantum regime. Then how can we choose the arrow of time? Isn't it so strange if we can find a definite direction of the time even in this quantum gravitational regime?

Perhaps, this conceptual tension might be explained if we introduce the Hartle-Hawking wave function \cite{Brahma:2018elv}. In order to compute the Euclidean Lagrangian $L_{\mathrm{E}}$ from the loop quantum gravity modified Euclidean Hamiltonian $H_{\mathrm{E}}$, we follow the relation:
\begin{eqnarray}
L_{\mathrm{E}} = p_{a}\dot{a} - H_{\mathrm{E}},
\end{eqnarray}
where $p_{a}$ is the canonical momentum of $a$. However, due to the Hamiltonian constraint, $H_{\mathrm{E}} = 0$ in the on-shell level description. Therefore, the Euclidean action is simply
\begin{eqnarray}
S_{\mathrm{E}} = - \frac{3\pi}{2} \int a \dot{a}^{2} d\tau = - \frac{3\pi}{2} \int_{0}^{a_{\mathrm{min}}} a\sqrt{|\mathcal{V}(a)|} da.
\end{eqnarray}

Interestingly, in the Euclidean signatures, as $a$ approaches to zero, $\mathcal{V}(a)$ approaches to zero, too. This implies that the instanton explains the infinitely stretched solution as a function of $\tau$, although the probability is well-defined \cite{Brahma:2018kkr}. Except for this, the interpretation is the same as that of the usual Hartle-Hawking wave function. Therefore, near the quantum bouncing point, the bouncing interpretation is not the unique way to understand; a universe can be created from nothing (Fig.~\ref{fig:pots}, \cite{Brahma:2018elv}). Also, in some parameter regimes, a Euclidean wormhole solution is possible. Then around the quantum bounding point, there is an ambiguity to define the arrow of time; either, a contracting phase bounces to an expanding phase, or two expanding universes are created via a Euclidean wormhole solution.

\subsection{Fuzzy instantons in anti-de Sitter space}

Finally, we report some discussions about the anti-de Sitter space. Let us consider the following potential:
\begin{eqnarray}
V(\Phi) = V_{0} \left( -1 - \frac{1}{2} \mu^{2} \Phi^{2} + \lambda \Phi^{4} \right).
\end{eqnarray}
In the Euclidean domain, it is not surprising to have a complex-valued solution. So, we consider the pure imaginary field: $\Phi \rightarrow i\phi$. Then the potential is effectively
\begin{eqnarray}
U(\phi) = V_{0} \left( -1 + \frac{1}{2} \mu^{2} \phi^{2} + \lambda \phi^{4} \right),
\end{eqnarray}
while the kinetic term has an opposite sign. Therefore, it is possible to find a solution that the scalar field asymptotically approaches to zero, while at the center, there may exist a throat \cite{Kang:2017jmq}.

\end{paracol}
\begin{figure}
\widefigure
\includegraphics[scale=0.3]{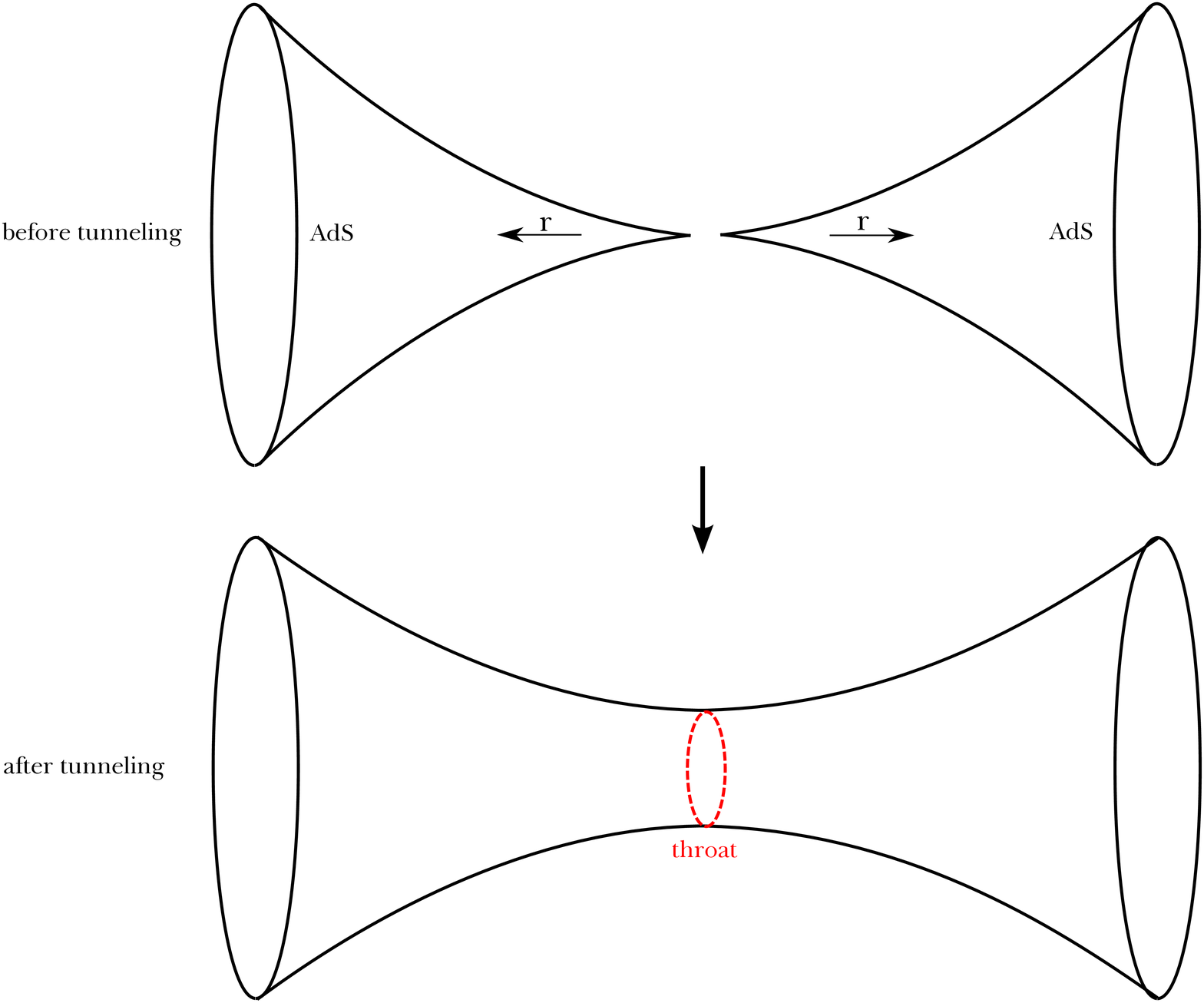}
\caption{\label{fig:concept}Conceptual picture of tunneling in the anti-de Sitter space \cite{Kang:2017jmq}.}
\end{figure}  
\begin{paracol}{2}
\linenumbers
\switchcolumn

One potential issue is whether the Euclidean action is well-defined or not. Since the volume of the Euclidean anti-de Sitter is infinite, the Euclidean action itself is infinite. However, by subtracting to the pure anti-de Sitter background, one may obtain a finite action difference. The sufficient condition to obtain the finite action difference is that the field should sufficiently quickly approach zero near the infinity \cite{Kanno:2012ht}. This can be achieved if we tune the shape of potential \cite{Kang:2017jmq}.

If this finite action difference is allowed, then we can believe that this instanton explains, after the Wick-rotation, a tunneling from two separate anti-de Sitter spaces to a connected anti-de Sitter wormhole (Fig.~\ref{fig:concept}, \cite{Kang:2017jmq}). This solution satisfies the classicality at the time-like infinity. One potential question is, whether any effects from the fuzzy core of the solution can reach future infinity or not, but in principle, this solution is well embedded in the Euclidean path integral formalism. Therefore, if we extend this technique, we can very easily extend to the anti-de Sitter fuzzy Euclidean wormholes in a black hole background; after the Wick-rotation, this explains a Lorentzian (probably unstable) wormhole in the anti-de Sitter space. At once this structure exists, this can cause conceptual trouble with the ER=EPR conjecture \cite{Maldacena:2013xja}. We leave these interesting connections to the information loss paradox for future research topics \cite{Chen:2016nvj,Chen:2020nag}.

\section{Future perspectives}

In this paper, we have discussed various aspects of fuzzy instantons of the Hartle-Hawking wave function.

First, we studied the slow-roll inflation models. Due to the fuzzy instanton analysis and the classicality condition, the universe should experience a little bit of inflation. However, the amount is not sufficient. We need several ways to rescue the Hartle-Hawking wave function to fit the observations. Perhaps, the most natural way is to introduce a massive field and impose the classicalization of all matter fields.

Second, we studied when the slow-roll conditions break down. This is very natural in the point of view of the UV-completed theory, e.g., string theory. However, a classical universe can be created even with these cases; we named these new types of solutions HSY instantons.

Third, we extended to diverge situations. For example, the Euclidean path integral formalism does not necessarily imply that the boundary is only one; in principle, it can have two boundaries. This case is related to the fuzzy Euclidean wormholes. Fuzzy Euclidean wormholes in the de Sitter space can explain the preference of the concave inflaton potential. Fuzzy Euclidean wormholes in the anti-de Sitter space may be related to the information loss paradox as well as a possible criticism of the ER=EPR conjecture. In addition to them, the no-boundary wave function can be applied for string-inspired models or loop quantum cosmology models. These models also allow Euclidean wormhole solutions. 

There may be several interesting research directions.
\begin{itemize}
\item[1.] Traditionally, we assumed landscape and considered slow-roll inflation models. However, in recent discussions, models which are consistent with the swampland criteria might be more interesting. The Hartle-Hawking wave function is definitely useful for both problems. However, it might be more interesting to provide possible observational consequences \cite{Chen:2017aes,Chen:2019mbu} that may reveal some issues related to the swampland criteria.
\item[2.] Fuzzy Euclidean wormholes can be realized in various systems, but it might be complicated to apply the technique beyond Einstein gravity. This may include the Gauss-Bonnet-dilaton gravity model or the loop quantum cosmological model. Some fuzzy extensions of oscillating instantons are also interesting \cite{Lee:2012qv}. In any case, this will be a challenging topic.
\item[3.] Fuzzy instantons in anti-de Sitter background or black hole background is also an interesting topic. This issue may also cover some topics of Hawking radiation \cite{Chen:2018aij} as well as the information loss problem \cite{Sasaki:2014spa}. However, it is also fair to say that it is not easy to impose the classicality condition at future infinity if the symmetry is lesser than the $O(4)$-symmetry. The generalization of the dynamical instantons in the spherical symmetry will be an important topic.
\end{itemize}

In conclusion, the Euclidean path integral is well approximated by instantons. If the instantons are dynamical, then they must be fuzzy or complexified. The investigation of dynamical wormholes is a challenging and fruitful future research topic. This is necessarily related to the study of fuzzy instantons, not only in the context of cosmology but also in black hole physics. We leave these fascinating research topics for the future.

\acknowledgments{This work is supported by the National Research Foundation of Korea (Grant no.:2021R1C1C1008622, 2021R1A4A5031460).}

\end{paracol}
\reftitle{References}


\begin{thebibliography}{999}

\bibitem{Hawking:1969sw}
S.~W.~Hawking and R.~Penrose,
Proc. Roy. Soc. Lond. A \textbf{314} (1970), 529-548.

\bibitem{DeWitt:1967yk}
B.~S.~DeWitt,
Phys. Rev. \textbf{160} (1967), 1113-1148.

\bibitem{Vilenkin:1987kf}
A.~Vilenkin,
Phys. Rev. D \textbf{37} (1988), 888.

\bibitem{Halliwell:1984eu}
J.~J.~Halliwell and S.~W.~Hawking,
Phys. Rev. D \textbf{31} (1985), 1777.

\bibitem{Hartle:1983ai}
J.~B.~Hartle and S.~W.~Hawking,
Phys. Rev. D \textbf{28} (1983), 2960-2975.

\bibitem{Halliwell:1989dy}
J.~J.~Halliwell and J.~B.~Hartle,
Phys. Rev. D \textbf{41} (1990), 1815.

\bibitem{Hartle:2008ng}
J.~B.~Hartle, S.~W.~Hawking and T.~Hertog,
Phys. Rev. D \textbf{77} (2008), 123537
[arXiv:0803.1663 [hep-th]].

\bibitem{Hartle:2007gi}
J.~B.~Hartle, S.~W.~Hawking and T.~Hertog,
Phys. Rev. Lett. \textbf{100} (2008), 201301
[arXiv:0711.4630 [hep-th]].

\bibitem{Hwang:2012mf}
D.~Hwang, B.~H.~Lee, E.~D.~Stewart, D.~Yeom and H.~Zoe,
Phys. Rev. D \textbf{87} (2013) no.6, 063502
[arXiv:1208.6563 [gr-qc]].

\bibitem{Hwang:2014vba}
D.~Hwang, S.~A.~Kim and D.~Yeom,
Class. Quant. Grav. \textbf{32} (2015) no.11, 115006
[arXiv:1404.2800 [gr-qc]].

\bibitem{Chen:2015ria}
P.~Chen, T.~Qiu and D.~Yeom,
Eur. Phys. J. C \textbf{76} (2016) no.2, 91
[arXiv:1503.08709 [gr-qc]].

\bibitem{Hwang:2012bd}
D.~Hwang, S.~A.~Kim, B.~H.~Lee, H.~Sahlmann and D.~Yeom,
Class. Quant. Grav. \textbf{30} (2013), 165016
[arXiv:1207.0359 [gr-qc]].

\bibitem{Feldbrugge:2017kzv}
J.~Feldbrugge, J.~L.~Lehners and N.~Turok,
Phys. Rev. D \textbf{95} (2017) no.10, 103508
[arXiv:1703.02076 [hep-th]].

\bibitem{DiazDorronsoro:2017hti}
J.~Diaz Dorronsoro, J.~J.~Halliwell, J.~B.~Hartle, T.~Hertog and O.~Janssen,
Phys. Rev. D \textbf{96} (2017) no.4, 043505
[arXiv:1705.05340 [gr-qc]].

\bibitem{Hwang:2013nja}
D.~Hwang and D.~Yeom,
JCAP \textbf{06} (2014), 007
[arXiv:1311.6872 [gr-qc]].

\bibitem{Sasaki:2013aka}
M.~Sasaki, D.~Yeom and Y.~l.~Zhang,
Class. Quant. Grav. \textbf{30} (2013), 232001
[arXiv:1307.5948 [gr-qc]].

\bibitem{Zhang:2014wia}
Y.~l.~Zhang, M.~Sasaki and D.~Yeom,
JHEP \textbf{04} (2015), 016
[arXiv:1411.6769 [hep-th]].

\bibitem{Susskind:2003kw}
L.~Susskind,
[arXiv:hep-th/0302219 [hep-th]].

\bibitem{Obied:2018sgi}
G.~Obied, H.~Ooguri, L.~Spodyneiko and C.~Vafa,
[arXiv:1806.08362 [hep-th]].

\bibitem{Brahma:2020cpy}
S.~Brahma, R.~Brandenberger and D.~Yeom,
JCAP \textbf{10} (2020), 037
[arXiv:2002.02941 [hep-th]].

\bibitem{Hwang:2012zj}
D.~Hwang, B.~H.~Lee, H.~Sahlmann and D.~Yeom,
Class. Quant. Grav. \textbf{29} (2012), 175001
[arXiv:1203.0112 [gr-qc]].

\bibitem{Hwang:2011mp}
D.~Hwang, H.~Sahlmann and D.~Yeom,
Class. Quant. Grav. \textbf{29} (2012), 095005
[arXiv:1107.4653 [gr-qc]].

\bibitem{Chen:2016ask}
P.~Chen, Y.~C.~Hu and D.~Yeom,
JCAP \textbf{07} (2017), 001
[arXiv:1611.08468 [gr-qc]].

\bibitem{Chew:2020lkj}
X.~Y.~Chew, G.~Tumurtushaa and D.~Yeom,
Phys. Dark Univ. \textbf{32} (2021), 100811
[arXiv:2006.04344 [gr-qc]].

\bibitem{Brahma:2018elv}
S.~Brahma and D.~Yeom,
Phys. Rev. D \textbf{98} (2018) no.8, 083537
[arXiv:1808.01744 [gr-qc]].

\bibitem{Chen:2017qeh}
P.~Chen and D.~Yeom,
Eur. Phys. J. C \textbf{78} (2018) no.10, 863
[arXiv:1706.07784 [gr-qc]].

\bibitem{Chen:2019cmw}
P.~Chen, D.~Ro and D.~Yeom,
Phys. Dark Univ. \textbf{28} (2020), 100492
[arXiv:1904.00199 [gr-qc]].

\bibitem{Starobinsky:1980te}
A.~A.~Starobinsky,
Phys. Lett. B \textbf{91} (1980), 99-102.

\bibitem{Akrami:2018odb}
Y.~Akrami \textit{et al.} [Planck],
Astron. Astrophys. \textbf{641} (2020), A10
[arXiv:1807.06211 [astro-ph.CO]].

\bibitem{Kanti:2011jz}
P.~Kanti, B.~Kleihaus and J.~Kunz,
Phys. Rev. Lett. \textbf{107} (2011), 271101
[arXiv:1108.3003 [gr-qc]].

\bibitem{Tumurtushaa:2018agq}
G.~Tumurtushaa and D.~Yeom,
Eur. Phys. J. C \textbf{79} (2019) no.6, 488
[arXiv:1808.01103 [hep-th]].

\bibitem{Bojowald:2018xxu}
M.~Bojowald, S.~Brahma and D.~Yeom,
Phys. Rev. D \textbf{98} (2018) no.4, 046015
[arXiv:1803.01119 [gr-qc]].

\bibitem{Brahma:2018kkr}
S.~Brahma and D.~Yeom,
Universe \textbf{5} (2019) no.1, 22
[arXiv:1810.10211 [hep-th]].

\bibitem{Kang:2017jmq}
S.~Kang and D.~Yeom,
Phys. Rev. D \textbf{97} (2018) no.12, 124031
[arXiv:1703.07746 [gr-qc]].

\bibitem{Kanno:2012ht}
S.~Kanno, M.~Sasaki and J.~Soda,
Prog. Theor. Phys. \textbf{128} (2012), 213-226
[arXiv:1203.0612 [hep-th]].

\bibitem{Maldacena:2013xja}
J.~Maldacena and L.~Susskind,
Fortsch. Phys. \textbf{61} (2013), 781-811
[arXiv:1306.0533 [hep-th]].

\bibitem{Chen:2016nvj}
P.~Chen, C.~H.~Wu and D.~Yeom,
JCAP \textbf{06} (2017), 040
[arXiv:1608.08695 [hep-th]].

\bibitem{Chen:2020nag}
P.~Chen, M.~Sasaki and D.~Yeom,
Int. J. Mod. Phys. D \textbf{29} (2020) no.12, 2050086
[arXiv:2005.07011 [gr-qc]].

\bibitem{Chen:2017aes}
P.~Chen, Y.~H.~Lin and D.~Yeom,
Eur. Phys. J. C \textbf{78} (2018) no.11, 930
[arXiv:1707.01471 [gr-qc]].

\bibitem{Chen:2019mbu}
P.~Chen, H.~H.~Yeh and D.~Yeom,
Phys. Dark Univ. \textbf{27} (2020), 100435
[arXiv:1903.12045 [gr-qc]].

\bibitem{Lee:2012qv}
B.~H.~Lee, W.~Lee and D.~Yeom,
Int. J. Mod. Phys. A \textbf{28} (2013), 1350082
[arXiv:1206.7040 [hep-th]].

\bibitem{Chen:2018aij}
P.~Chen, M.~Sasaki and D.~Yeom,
Eur. Phys. J. C \textbf{79} (2019) no.7, 627
[arXiv:1806.03766 [hep-th]].

\bibitem{Sasaki:2014spa}
M.~Sasaki and D.~Yeom,
JHEP \textbf{12} (2014), 155
[arXiv:1404.1565 [hep-th]].




\end{thebibliography}
\end{document}